\documentclass{article}

\usepackage{PRIMEarxiv}

\usepackage[utf8]{inputenc} 
\usepackage[T1]{fontenc}    
\usepackage{hyperref}       
\usepackage{url}            
\usepackage{booktabs}       
\usepackage{amsfonts}       
\usepackage{nicefrac}       
\usepackage{microtype}      
\usepackage{lipsum}
\usepackage{fancyhdr}       
\usepackage{graphicx}       
\graphicspath{{media/}}     
\usepackage{caption}
\usepackage{subcaption}
\usepackage{algorithm}
\usepackage{algpseudocode}

\usepackage{todonotes}
\usepackage{amsmath}
\usepackage{multirow}

\usepackage{lmodern}

\usepackage{listings}

\colorlet{punct}{red!60!black}
\definecolor{background}{HTML}{EEEEEE}
\definecolor{delim}{RGB}{20,105,176}
\colorlet{numb}{magenta!60!black}

\lstdefinelanguage{json}{
    basicstyle=\normalfont\ttfamily,
    numbers=left,
    numberstyle=\scriptsize,
    stepnumber=1,
    numbersep=8pt,
    showstringspaces=false,
    breaklines=true,
    frame=lines,
    backgroundcolor=\color{background},
    literate=
     *{0}{{{\color{numb}0}}}{1}
      {1}{{{\color{numb}1}}}{1}
      {2}{{{\color{numb}2}}}{1}
      {3}{{{\color{numb}3}}}{1}
      {4}{{{\color{numb}4}}}{1}
      {5}{{{\color{numb}5}}}{1}
      {6}{{{\color{numb}6}}}{1}
      {7}{{{\color{numb}7}}}{1}
      {8}{{{\color{numb}8}}}{1}
      {9}{{{\color{numb}9}}}{1}
      {:}{{{\color{punct}{:}}}}{1}
      {,}{{{\color{punct}{,}}}}{1}
      {\{}{{{\color{delim}{\{}}}}{1}
      {\}}{{{\color{delim}{\}}}}}{1}
      {[}{{{\color{delim}{[}}}}{1}
      {]}{{{\color{delim}{]}}}}{1},
}

\title{Language Models are Spacecraft Operators
}

\author{
  Victor Rodriguez-Fernandez, Alejandro Carrasco \\
  Universidad Politécnica de Madrid \\
  Madrid 28038, Spain \\
  \texttt{\{victor.rfernandez@, alejandro.carrasco.aragon@alumnos.\}upm.es} \\
  \And
  Jason Cheng, Eli Scharf, Peng Mun Siew, Richard Linares \\
  Massachusetts Institute of Technology \\
  Cambridge, Massachusetts 02139, USA \\
  \texttt{\{jasonc75, escharf, siewpm, linaresr\}@mit.edu} \\
}

\begin{document}
\maketitle

\begin{abstract}
Recent trends are emerging in the use of Large Language Models (LLMs) as autonomous agents that take actions based on the content of the user text prompts. We intend to apply these concepts to the field of Guidance, Navigation, and Control in space, enabling LLMs to have a significant role in the decision-making process for autonomous satellite operations. As a first step towards this goal, we have developed a pure LLM-based solution for the Kerbal Space Program Differential Games (KSPDG) challenge, a public software design competition where participants create autonomous agents for maneuvering satellites involved in non-cooperative space operations, running on the KSP game engine. Our approach leverages prompt engineering, few-shot prompting, and fine-tuning techniques to create an effective LLM-based agent that ranked 2nd in the competition. To the best of our knowledge, this work pioneers the integration of LLM agents into space research. Code is available at this \href{https://github.com/ARCLab-MIT/kspdg}{url}.
\end{abstract}


\section{Introduction}
Large Language Models (LLMs) are, without a doubt, the last major breakthrough in the evolution of artificial intelligence systems. Since the release of ChatGPT \cite{chatgpt} at the end of 2022, we have seen a plethora of applications and use cases emerge across various industries. From generating human-like text to aiding in code completion, LLMs have significantly impacted the way we interact with technology and the possibilities of what AI can achieve.

In recent months, the use of LLMs is expanding beyond text-based applications to become \textit{language agents} capable of taking actions based on the context of the system in which they are integrated. By leveraging the contextual information available to them, LLMs can make informed decisions and perform tasks autonomously. Common examples include LLMs connecting to a web browser or an external API to give more accurate responses, but researchers are also applying this idea to the physical world, creating things like LLM-driven robot agents \cite{wang2023survey} and motion planners that generate driving trajectories for self-driving cars \cite{mao2023gpt}. This new way of creating autonomous agents intersects with the usage of  Reinforcement Learning (RL) algorithms, and provides a way to overcome some of its well-known limitations, such as the sample inefficiency, and the need for a well-defined reward function. Some recent studies have demonstrated how some powerful LLMs, such as GPT-4, can surpass state-of-the-art RL algorithms in complex games just through studying academics texts and reasoning \cite{wu2023spring}, executing sophisticated trajectories and achieving good zero-short performance.

This work is focused on the domain of space applications and the development of autonomous agents for guidance and control of spacecrafts. In this context, the creation of AI-based agents has mainly been tackled through RL in recent years, and we can find agents trained for different tasks such as sensor-tasking \cite{siew2022space} and planetary landing \cite{gaudet2020deep}. However, unlike other AI research areas, the space domain lacks of publicly available simulation environments, which are crucial for training AI agents in complex space operations and providing a standard benchmark for evaluating different AI and autonomous control methods. To address this issue, Allen et al. introduced \textit{SpaceGym} \cite{10115968}, a set non-cooperative game environments that are intended to spur development and act as proving grounds for autonomous and AI decision-makers in the orbital domain. Among the available environments in \textit{SpaceGym}, in this work we focus on the Kerbal Space Program Differential Games suite (KSPDG). KSPDG is a suite of differential games, such as pursuit-evasion scenarios, encoded within the Kerbal Space Program game engine \footnote{https://store.privatedivision.com/game/kerbal-space-program-2} and standardized with OpenAI Gym \cite{brockman2016openai} and PettingZoo \cite{terry2021pettingzoo} interfaces, facilitating the use of diverse AI techniques, including multi-agent reinforcement learning.

While KSPDG presents an innovative framework for testing AI and autonomous control methods in space applications, it is unsuitable for RL training, due to technical and non-technical reasons. On the one hand, the KSP engine, which underpins KSPDG, lacks the capacity for the parallel, accelerated, and headless operations essential for extensive faster-than-real-time RL training. On the other hand, the principled stance of KSPDG's creators to focus on evaluation rather than training emphasizes the need for a ``true test set" environment where overfitting is minimized, and the genuine and unbiased capabilities of AI agents are tested. This approach diverges from the typical RL methodology that relies on iterative training and fine-tuning of agents within a specific simulation environment.

\begin{figure}
  \centering
  \includegraphics[width=\textwidth]{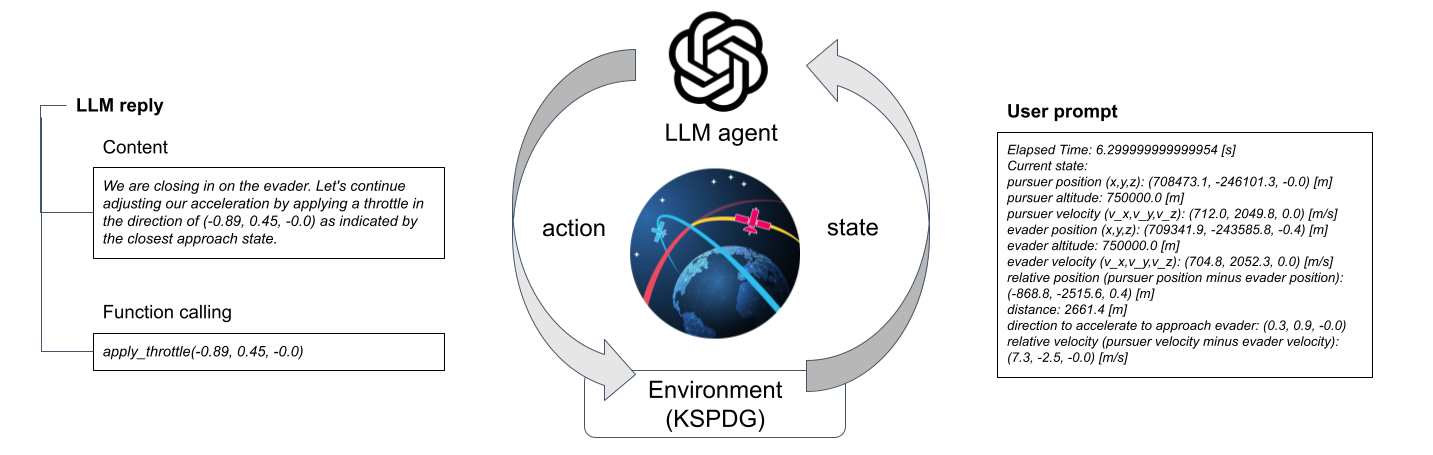}
  \caption{Overview of the proposed approach to use a LLM (e.g. ChatGPT) as an autonomous spacecraft operator that gets, as user prompt, the current status of the mission from the KSDPG simulation environment (i.e., the state or observation in the RL jargon), and replies with a reasoned action to carry out, expressed as a function calling with the specific throttle vector and the textual justification behind the action.}
  \label{fig:overview}
\end{figure}

To overcome the limitations of RL in creating autonomous agents for environments such as KSDPG, as well as for other space operations where numerous simulated data cannot be provided, we propose to adapt the current trend of LLM-based agents to develop an ``intelligent" operator that controls a spacecraft based on the real-time telemetry of the environment, using exclusively language as the input and output of the system. As depicted in Fig. \ref{fig:overview}, we design the classic RL loop by interfacing the simulation environment (KSDPG) with a LLM, transforming the real-time observations (or state) of the mission as textual user prompts that are fed to the model. The LLM then processes the prompt and replies with an action that will be plugged in KSDPG to control the spacecraft. In this work, we use GPT 3.5 as LLM, and the OpenAI API as the endpoint to interact with it. This API provides a way to describe functions and have the model intelligently choose to output a ``function calling" object as part of the reply, containing arguments to call the function in case it is needed. We experimented with different strategies to prompt the model with the KSPDG observations, as well as to fine-tune it with example data collected from human gameplays. Our agent was ranked 2nd in the KSPDG challenge \footnote{https://www.ll.mit.edu/conferences-events/2024/01/kerbal-space-program-differential-game-challenge}, and was presented via a live demonstration during a special session at AIAA SciTech in January 2024.

The rest of the paper is structured as follows: Section \ref{sec:backgrounds} provides a comprehensive background, introducing the reader to the Kerbal Space Program Differential Games (KSPDG) environment, as well as LLMs and the OpenAI GPT API, which form the foundation of our research. In the subsequent sections, we discuss the methodologies employed in our approach, starting with prompt engineering and observation augmentation in Section 3.1, which outlines how we optimized LLM performance for the KSPDG environment. Section 3.2 explores the application of few-shot prompting to improve LLM responses, while Section 3.3 details our fine-tuning strategies to enhance model efficiency and accuracy. Each of these sections not only presents our approach but also illustrates the impact of these techniques through experimental results. Finally, the paper concludes in Section 4 with a summary of our findings, the implications of integrating LLMs into spacecraft operations, and potential directions for future research.

\section{Backgrounds}
\label{sec:backgrounds}


\subsection{KSPDG: Kerbal Space Program Differential Games}
Kerbal Space Program (KSP) is a popular space exploration computer video game. The game allows players to explore a fictional space system. KSP is known for its realistic physics, orbital mechanics, and rocket design. The game gives players great freedom in designing their spacecraft, space missions, and space exploration.

To bridge the gap between research in autonomous agents and space exploration simulations, the Massachusetts Institute of Technology’s Lincoln Laboratories (MIT-LL) developed the space-gym library \cite{10115968}. Building on the foundation of the widely-used OpenAI Gym API \cite{brockman2016openai}, space-gym extends its functionality to the domain of KSP, offering a familiar RL interface tailored specifically for spacecraft operations. 

The MIT-LL hosted the \textit{Kerbal Space Program Differential Game challenge} (KSPDG)\cite{kspdg}. This challenge tasked participants with developing autonomous agents and Artificial Intelligence algorithms for maneuvering satellites in non-cooperative space operations. Non-cooperative space operations include pursuing an evasive satellite and multi-satellite proximity operations. Participants' autonomous agents were evaluated on criteria such as time to completion of missions, fuel consumption of agents, relative distance to the evader satellite, and minimum relative position-velocity product. 

In spacegym-KSPDG, agent’s throttles are input to the KSP game engine as a list of three integers between negative one and one. For example, three positive ones correspond to a full forward, right, and down throttle, whereas three negative ones correspond to backward, left, and up. A fourth number is appended to the list corresponding to the duration the throttles should be applied. Throttle vectors are expressed with respect to the spacecraft's reference frame. 

Once the execution time of a throttle action is over, spacegym-KSPDG returns the current state of the mission to the agent as an observation vector. In the case of the Pursuer-Evader scenario, the vector contains the following information: mission time (s), vehicle mass (kg), vehicle fuel (kg), pursuer x-position, pursuer y-position, pursuer z-position, pursuer x-velocity (m/s), pursuer y-velocity (m/s), pursuer z-velocity (m/s), evader x-position, evader y-position, evader z-position, evader x-velocity (m/s), evader y-velocity (m/s), evader z-velocity (m/s). 

\begin{figure} [h!]
\centering
\begin{subfigure}{.5\textwidth}
  \centering
  \includegraphics[width=0.75\linewidth]{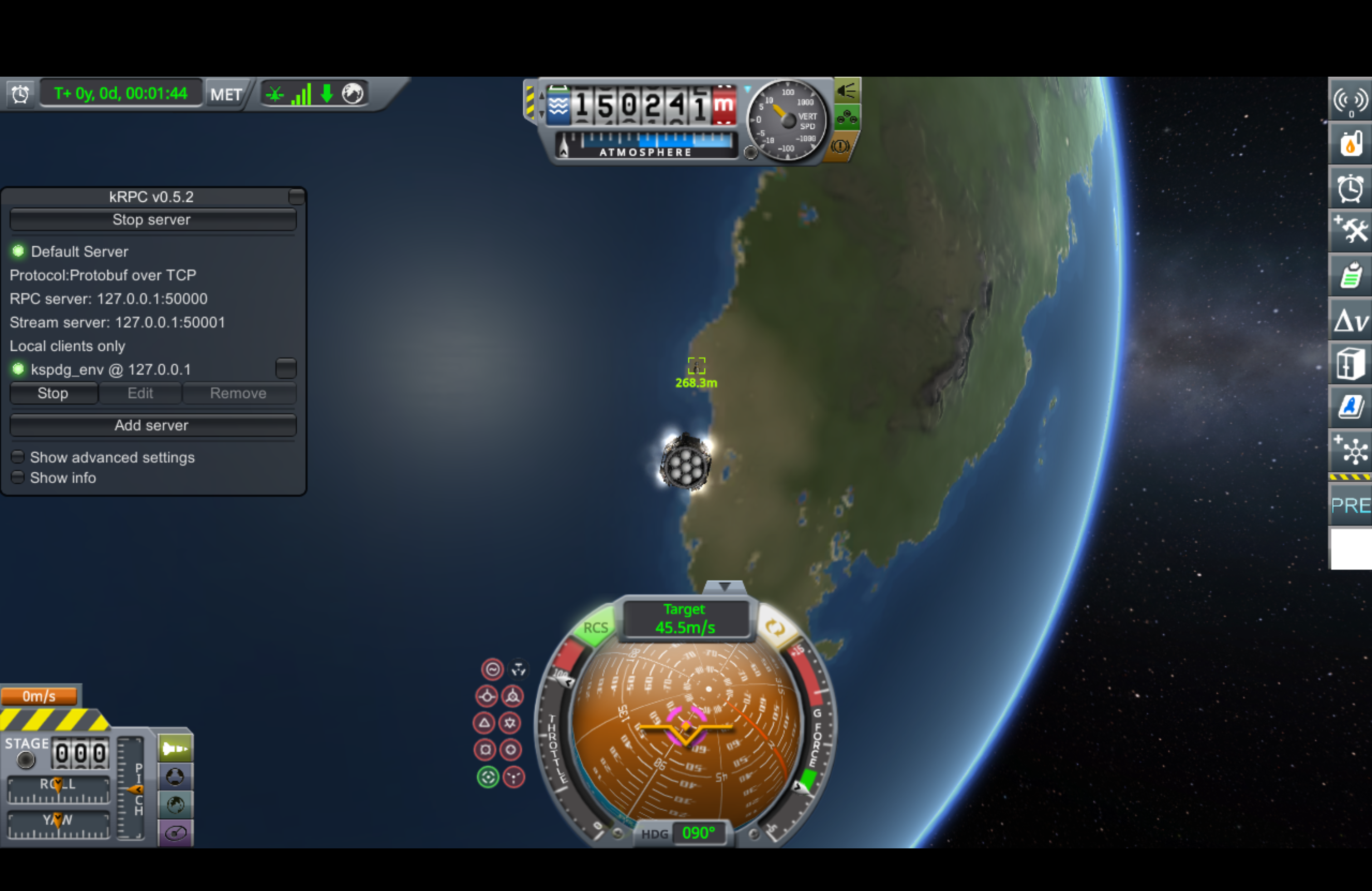}
  \caption{Pursuer-Evader Scenario}
  \label{fig:sub1}
\end{subfigure}%
\begin{subfigure}{.5\textwidth}
  \centering
  \includegraphics[width=0.75\linewidth]{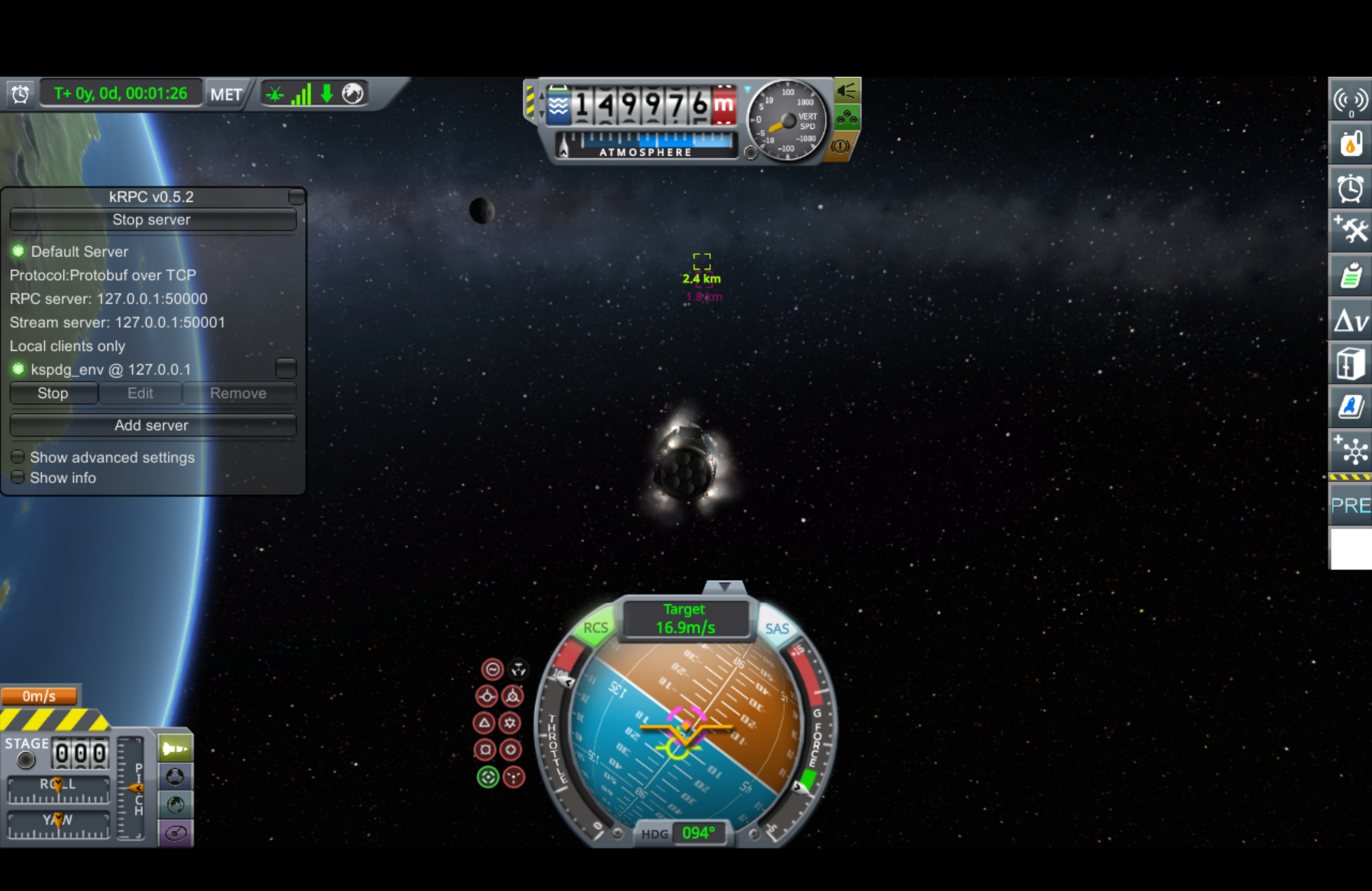}
  \caption{Lady-Bandit-Guard Scenario}
  \label{fig:sub2}
\end{subfigure}
\caption{Kerbal Space Program Differential Games. In this work we will focus on the Pursuer-Evader scenario.}
\label{fig:test}
\end{figure}

KSPDG consists of three main categories of scenarios (See Fig. \ref{fig:test}): Pursuer-Evader, multi-agent target guarding, and 1-v-1 sun blocking problem. In all three categories, the action and observation spaces remained the same. 
\begin{itemize}
    \item Pursuer-evader Scenarios (scenarios E1, E2, E3, E4): participants built autonomous agents to control the pursuer. The main objective was to minimize the distance between the pursuer and the evader. For different scenarios in the Pursuer-Evader game, the evader’s initial orbit remained constant across all scenarios while the pursuer’s initial orbit varied. The pursuer and evader have identical vehicle parameters across all games in this scenario. Participants were evaluated on metrics such as distance between pursuer and evader, speed at closest approach, pursuer fuel usage, and time elapsed. the main differences between the E1, E2, E3, and E4 scenarios lie in the evasive maneuvers implemented by the Evader vessel: 
    \begin{itemize}
        \item In E1, no evasive maneuvers are defined.
        \item E2 introduces random evasive maneuvers, where the Evader vessel randomly adjusts its thrust and attitude for short durations when the Pursuer is within a certain range.
        \item E3 employs a more structured evasive strategy, where the Evader vessel activates full thrust to escape when the Pursuer is within a specified distance threshold.
        \item E4 has the evader vessel perform a simple prograde evasion by aligning itself with the orbital velocity vector and applying a constant throttle.
    \end{itemize}

    \item  Target guarding Scenarios (scnearios LBG1, LBG2): these scenarios consisted of 1-v-2 target guarding. These scenarios are also called Lady-Bandit-Guard scenarios \cite{lbg}. Participants designed autonomous agents to control the bandit. The main objective was to minimize the distance between the bandit and the lady while also maximizing the distance between the guard and the bandit. A scripted bot controlled the lady and the guard, but the bot's policy varies between environments within the group. The bandit and guard have identical vehicle capabilities in each scenario. The Lady’s vehicle capabilities varied for each scenario. The lady’s initial orbit was kept constant across all scenarios. The bandit’s and guard’s initial orbits varied across environments. The scoring function was a sum of the closest lady-bandit approach distance squared and the inverse of the closest bandit-guard approach.
    \item Sun-blocking Scenarios (scenarios SB1, SB2, SB3, SB4, SB5): The 1-v-1 sun-blocking problem tasked participants with building an autonomous agent to arrive at a co-linear point between the evader and the sun. A scripted bot controlled the evader. The bots policy varied between scenarios in this category. The pursuer and evader had identical vehicle parameters across all scenarios. The evader’s initial orbit was held constant across all scenarios, while the pursuer’s initial orbit varied. 
\end{itemize}

\subsection{Large Language Models}
Natural language processing (NLP) aims at enhancing human-machine interface by improving how machines can understand and manipulate natural language text or speech to complete a task, or even communicate with humans using natural language. Over the last decades, large strides have been achieved in the field of NLP and this forms the basis and foundation for the study of Large Language Models (LLM). LLMs are a specific class of NLP models that are trained to predict the next token (subword) in a sentence. They typically consist of billions or even trillions of parameters, and are trained on massive datasets, allowing them to learn the intricacies of natural language from a simple task such as one-token ahead prediction. Some examples of popular LLMs include OpenAI's ChatGPT \cite{achiam2023gpt}, Google's Gemini \cite{team2023gemini}, and Microsoft's Copilot \footnote{MS copilot: \url{https://copilot.microsoft.com/}}. These LLMs have revolutionized how humans interact with AI and machines.

Google's Gemini leverages Google's pre-trained Language Model for Dialogue Applications (LaMDA) \cite{thoppilan2022lamda} and further refines the model for conversational interactions, whereas both OpenAI's ChatGPT and Microsoft's Bing Chat are built upon OpenAI's Generative Pre-trained Transformer (GPT) language model.
All these LLM models share a common origin; the transformer network. The transformer network revolutionized the NLP landscapes by introducing the self-attention mechanism, which allows the model to process a whole data sequence simultaneously, and provides great scaling and parallelization capabilities on GPUs \cite{vaswani2017attention}. 

LLM models were originally designed to perform language-related tasks with a high degree of accuracy, reasoning abilities, and contextual understanding. Notably, these capabilities were extended to include processing and comprehension of images in the latest multimodal GPT model (GPT-4v) \cite{gpt4v}. With the popularization and release of API access to LLM models, various novel researches have been conducted to leverage LLM models across diverse domains, ranging from education and healthcare to robotics and automation \cite{yang2023llmgrounder, mao2023gpt, xu2023drivegpt4}. 
\cite{yang2023llmgrounder} proposed a zero-shot, open-vocabulary, LLM-based 3D visual grounding pipeline that utilizes LLM to parse complex natural language queries into commands/sub-queries for visual grounder tools for object localization within the 3D space. Meanwhile, \cite{mao2023gpt} demonstrated the versatility of LLM models as high-level motion planners for autonomous vehicles. They constructed specialized LLM input prompts based on raw observations, guiding the model to generate valid trajectories accompanied by reasoning in natural language. On the other hand, \cite{xu2023drivegpt4} explored the usage of multimodal LLM to interpret, provide reasoning, and predict future actions of autonomous vehicles using video and past control signals.

\subsubsection{OpenAI's ChatGPT API}
OpenAI's ChatGPT offers a so-called \textit{completions} API endpoint, granting users access to the GPT-3.5 and GPT-4 models for custom codes or software. This feature allows code developers and enthusiasts to easily leverage the power of LLMs over a diverse set of applications.
At the time of writing, the completions API has been designated as legacy and has been succeeded by the \textit{chat completions} API endpoint. It is important to note that the work detailed in this paper was conducted using the legacy completions API endpoint.

Other than the completions API endpoint, the function calling capabilities of ChatGPT were also leveraged for this work. Function calling allows programmers to customize the output of the LLM to meet specific functional requirements.
Through a single completions API call, users can define a library of custom functions, prompting the LLM to generate a JSON object containing arguments to call one or more specified functions depending on the scenario.
Function calls within the ChatGPT API enable developers to provide context, utilize predefined functions, guide the model's behavior, and extract relevant information from the generated responses rapidly. Listing \ref{lst:function_call} shows an example of the schema of a function ``apply\_throttle", ready to be used in ChatGPT's completions API endpoint.

\begin{lstlisting}[language=json,caption={Example of a function schema for the function ``apply\_throttle", used in this work. Adding this function schema to the \textit{functions} field of OpenAI's completions API enables function calling in the model response.},label={lst:function_call}]
  functions: [
    {
      "name": "apply_throttle",
      "description": "Move the pursuer spacecraft with a (x,y,z) throttle.",
      "parameters": {
        "type": "object",
        "properties": {
          "throttle": {
            "type": "array",
            "items": {
              "type": "number",
              "minimum": -1.0,
              "maximum": 1.0
            },
            "description": "An array of three floats, each between -1.0 and 1.0, that specify the (x,y,z) throttle values."
          }
        },
        "required": ["throttle"]
      }
    }
  ]
\end{lstlisting}


Finally, system prompting played a key role in guiding the behavior of the LLM. System prompting involves giving high-level instructions to the LLM, providing background context for the task. These instructions are passed into the API query every time the developer makes a call to the LLM API. By using system prompts correctly, developers can begin to fine-tune the LLM's response to align with KSPDG's goals. Leveraging system prompts showcases the adaptability of LLMs to complex scenarios such as the KSPDG. 

\section{Employed strategies}

\subsection{Prompt engineering and observation augmentation}
To optimize the performance of our LLM, we employ prompt engineering. Prompt engineering involves designing the input text to be more easily understood by a language model. First, we employ basic prompt engineering, where we change the wording and structure of our system and user prompts. We perform prompt engineering in a systematic way: we generate multiple variations on our current prompting, run simulations in KSPDG, and pick the prompt that achieves the best results. We find that the optimal prompts provide a concise explanation of KSP in the system prompt and give periodic observations in the user prompt, wrapped in explanatory text. We provide an example of our user prompt in Fig. \ref{fig:llm_prompt}.

However, we notice a few shortcomings that are not fixed with basic prompt engineering. One example is the LLM's poor ability to reason with the large numbers in our observation space. The observations include $(r_P, v_P, r_E, v_E)$, where $||r_P||, ||r_E|| \approx 7.5 * 10^5$, and $||v_P||, ||v_E|| \approx 2.0 * 10^3$. We find that the LLM is often inaccurate when performing arithmetic with large numbers, especially when they differ in magnitude.

Thus, another approach we took was to augment the observation space. We take inspiration from Retrieval Augmented Generation (RAG) \cite{lewis2021retrievalaugmented}, in which an LLM is provided with helper knowledge bases that it can call to aid its response. Well-known examples include GPT4's ability to use Bing search when formulating its response. Similar to RAG, we supply our LLM with additional observations that we calculate from the original observations. However, since RAG would add too much latency to our agent, we instead choose observations that are useful in all situations and concatenate them to every prompt. This skips the need for the LLM to call the retrieval function because we provide the results of what it might need in the initial prompt.

For the Pursuer-Evader scenario, we supply these additional observations:
\begin{itemize}
    \item relative position: \( r_{\text{rel}} = r_E - r_P \),
    \item distance to evader: \( d = \left\lVert r_{\text{rel}} \right\rVert \),
    \item relative velocity: \( v_{\text{rel}} = v_E - v_P \),
    \item direction of evader: \( u = \frac{r_{\text{rel}}}{\left\lVert r_{\text{rel}} \right\rVert} \),
    \item simulated ETA of closest approach: \( t_1 \), and
    \item direction of evader at the simulated closest approach: \( u_1 \).
\end{itemize}

We wrap each observation in explanatory text. An example of a full prompt we may use is shown in \ref{fig:llm_prompt}. To calculate $t_1$ and $u_1$, we solve the spacecraft's position at a future point in time with numerical integration. We repeat this at periodic intervals, e.g. every second, and calculate the state with the closest distance. From this state, we then derive the ETA and direction of the evader.

Augmenting the prompt with these observations significantly improves the agent's performance. We find that supplying these observations, and especially the direction observations, $u$ and $u_1$, significantly increases the LLM's capability to accelerate in the correct direction. One reason is that the LLM is often inaccurate when performing the calculations to get the direction $u = \frac{r_{\text{rel}}}{\left\lVert r_{\text{rel}} \right\rVert}$, or sometimes skips this altogether. This can be verified by checking the LLM's thought process in the response and noticing its inaccurate numerical calculations.

\begin{figure}[ht]
\centering
\fbox{
\begin{minipage}{.9\textwidth}
\texttt{Elapsed Time: 112.7799999999947 [s]\\
Current state:\\
pursuer position (x,y,z): (750044.3, -18124.4, 1.0) [m]\\
evader position (x,y,z): (749797.9, -17108.9, -0.0) [m]\\
relative position (pursuer position minus evader position): (246.4, -1015.5, 1.0) [m]\\
distance: 1044.9 [m]\\
direction to accelerate to approach evader: (-0.24, 0.97, -0.0)\\
relative velocity (pursuer velocity minus evader velocity): (-0.41, 27.55, 0.02) [m/s]\\
If your spacecraft and the evader both stop accelerating, this will be the simulated closest approach (happens at time 37s):\\
pursuer position (x,y,z): (747565.2, 63112.5, 2.0) [m]\\
evader position (x,y,z): (747331.4, 63104.8, 0.1) [m]\\
relative position at closest approach (pursuer position minus evader position): (233.8, 7.7, 1.8) [m]\\
distance at closest approach: 234.0 [m]\\
direction to accelerate to approach evader at closest approach: (-1.0, -0.03, -0.01)\\
relative velocity (pursuer velocity minus evader velocity) at closest approach: (-0.27, 27.71, 0.02) [m/s]}
\end{minipage}
}
\caption{Example user prompt to the LLM, showing prompt augmentation}
\label{fig:llm_prompt}
\end{figure}

\subsection{Few-shot prompting}
We notice that the LLM often fails to perform the function call in its first response, and this causes a negative chain reaction where the next responses also fail to perform the function call. To mitigate this, we manually write the first response, similar to traditional one-shot prompting technique \cite{brown2020language}. In this response, we manually write some reasoning that we want the LLM to take when deciding its acceleration and include a valid function call. We append this response to the conversation history. We observe that this greatly reduces the chance of erroneous function calls and improves the LLM's reasoning in later responses.

We notice that the LLM often copies the action taken in the first response in later responses: if the first response accelerates straight in the direction of the evader, subsequent responses will often do this as well. This hinders the ability of the LLM to change direction to intercept the evader correctly and results in bad average performance. To mitigate this, we investigate the effect of using further manual responses, each with a hard-coded trigger condition. The final list of responses we use, and their corresponding trigger conditions, is:

\begin{enumerate}
    \item In the first response, accelerate in $u$,
    \item When $t_1<60$ for the first time, accelerate in $u_1$ to adjust our intercept, and
    \item When $t_1<30$ for the first time, accelerate in $0.6*u_1-0.4*v_{rel}$ to continue adjusting our intercept while reducing the relative velocity.
\end{enumerate}

When we detect the trigger condition (e.g. $t_1$ dropping below 60 seconds for the first time), we append a manual response to the conversation history. For the manual response, we write reasoning in the form of text that we hope the LLM would take. We observe that subsequent calls to the LLM become improved by this manual prompting.

\begin{figure}[ht]
\centering
\fbox{
\begin{minipage}{.9\textwidth}
\texttt{We are getting close to the evader. We need to stop accelerating towards the current evader position. Instead, let's intercept the evader by accelerating in the direction of what the closest approach state tells us, which is (-0.75, 0.66, -0.0).\\
throttles: [-0.75, 0.66, -0.0]}
\end{minipage}
}
\caption{Example manual response for few-shot prompting}
\label{fig:manual_response}
\end{figure}

\begin{table}[ht]
\centering
\begin{tabular}{l|ccc}
\hline
\textbf{Method} & \textbf{Best Distance} & \textbf{Average Distance} & \textbf{Failure Rate} \\ \hline
Naive & 225.0 & 225.0 & - \\
PPO & >1643 & 2346 & - \\
iLQGames & >53.31 & 60.99 & - \\
Lambert-MPC & >18.24 & \textbf{47.94} & - \\
\hline
LLM & 292.5 & 328.2 & 30.0\% \\
w/ response 1 & 19.0 & 181.8 & 16.7\% \\
w/ response 1 \& 2 & \textbf{12.6} & \textbf{131.5} & \textbf{15.8}\% \\ \hline
\end{tabular}
\caption{Performance of LLM with varying manual prompting, measured as the closest distance achieved to the evader throughout the duration of an episode in the Pursuer-Evader game. The best distance scores for SpaceGym agents are not given; instead, we use a lower bound computed using the data about the mean distance, standard deviations, and number of trials given in SpaceGym.}
\label{tab:table_1_label}
\end{table}

Table 1 summarizes the performance of our agent with varying uses of manual responses. Erroneous function calling causes the environment to terminate or the agent to freeze; these are considered failures. Over all successful runs, we calculate the best closest approach distance and average closest approach distance. We compare these results against the agents used in SpaceGym \cite{10115968}. Specifically, we compare them against a naive agent, a Lambert model predictive control (Lambert-MPC) agent, an iLQGames agent \cite{fridovichkeil2020efficient}, and a Proximal Policy Optimization (PPO) agent, as benchmarked in the SpaceGym paper \cite{10115968}. We refer the reader to SpaceGym to read more about these agents. Response 3 is not included because we use distance as our comparison metric and do not consider relative velocity, as the closest approach distance is the largest factor that contributes to the score and is most easily interpreted. As shown in the table, adding more manual prompting greatly increases the performance of our agent. Additionally, it reduces the variance between runs. Only the domain-specific agents, the iLQGames agent and the Lambert-MPC agent, are able to achieve a closer average approach than our model. Still, neither are able to set the closest approach score, showing the effectiveness of large language models in this task.

To maximize the score, the given prompt proved effective in surpassing leaderboard scores, albeit at the cost of reduced generalization. Accordingly, we developed a new prompt aimed at not only efficiently accomplishing this task but also enhancing generalization. The most known technique that we used for creating this new prompt is called Chain of Thought (CoT) \cite{wei2023chain} which has been proved effective for eliciting reasoning in a model. CoT is a reasoning process where intermediate steps or conclusions are sequentially connected to arrive at a final answer or solution.

Utilizing the Chain of Thought (CoT) approach, the quality of the output is heavily reliant on the model's capacity for reasoning. Therefore, we employed the most advanced version available for GPT-3.5 model\footnote{The specific model version used in this CoT study is gpt-3.5-turbo-0125.}, to leverage its enhanced reasoning capabilities while maintaining the response speed characteristic of GPT-3.5. The prompts we used are as follows:

\begin{figure}[h]
\centering
\fbox{
\begin{minipage}{.9\textwidth}
\texttt{You operate as an autonomous agent controlling a pursuit spacecraft. Your goal is to apply
throttles to capture the evader given the positions and velocities of the pursuer and evader in celestial body
reference frame and the direction of pursuer's velocity relative to evader or prograde. Throttles must be given
in your vessel reference frame wherein the x axis points to the right, the y axis points towards the target and
the z axis points upwards. The maximum throttle is 1. Reason step-by-step.
After reasoning call the perform\_action function.}
\end{minipage}
}
\caption{System prompt used for eliciting Chain of Thought in this model}
\label{fig:system_prompt-CoT}
\end{figure}

\begin{figure}[H]
\centering
\fbox{
\begin{minipage}{.9\textwidth}

USER: Given these observations \{\textit{KSPDG environment observations from a past action}\}, 
what is the best throttle to capture evader?
\newline

ASSISTANT: The x coordinate of prograde is positive, indicating that evader is moving to the right. 
The y coordinate of prograde is negative, indicating that evader is approaching. 
The z coordinate of prograde is negative, indicating that evader is moving down. 
To capture the evader, we should move in the opposite direction in the x axis, 
towards the target in the y axis, and in the opposite direction in the z axis. 
This means we should apply throttles to move left, forward and up. Therefore we should call 
perform\_action(\{"ft": "forward", "rt": "left", "dt": "up"\}). \newline

Now answer the following question: \newline

Given these observations \{\textit{Current KSPDG environment observations}\}, 
what is the best throttle to capture evader?.
\end{minipage}
}
\caption{Sample of a user prompt selected from a log of gameplay interactions involving this specific model}
\label{fig:user_prompt_CoT}
\end{figure}

As depicted in Figure \ref{fig:system_prompt-CoT}, the provided system prompt introduces key terminologies such as \textit{prograde}, \textit{reference frame} and \textit{celestial body}. These terms are critical for understanding the context and analytical framework of the user prompt. Figure \ref{fig:user_prompt_CoT} illustrates an example of a user prompt. This example, derived from previous gameplay interactions, demonstrates the application of the previously mentioned terminologies in formulating a correct action.

\begin{table}[h]
\centering
\begin{tabular}{l|cc|cc|cc|cc|c}
\hline
\ref{sec:backgrounds}
\multirow{2}{*}{\textbf{Method}} & \multicolumn{2}{c|}{\textbf{E1 (m)}} & \multicolumn{2}{c|}{\textbf{E2 (m)}} & \multicolumn{2}{c|}{\textbf{E3 (m)}} & \multicolumn{2}{c|}{\textbf{E4 (m)}} & \multirow{2}{*}{\textbf{Failure Rate}} \\
\cline{2-9}
 & \textbf{Best} & \textbf{Avg} & \textbf{Best} & \textbf{Avg} & \textbf{Best} & \textbf{Avg} & \textbf{Best} & \textbf{Avg} & \\
\hline
Naive & 229.75 & 267.51 & 182.28 & 236.10 & 225.0 & 225.0 & 215.20 & 216.49 & - \\
Baseline LLM & 36.42 & 269.22 & 245.76 & 295.09 & 292.59 &  328.20 & 305.02 & 329.78 & 37.5\% \\
w/ CoT & \textbf{11.02} & \textbf{14.59} & \textbf{15.73} & \textbf{21.28} & \textbf{5.63} & \textbf{21.71} & \textbf{10.42} & \textbf{17.27} & 0.00\% \\ \hline
\end{tabular}
\caption{Performance (as closest distance) of the LLM agent with Chain of Thought (CoT) prompting across different environments of the Pursuer-Evader game. Each environment features different initial conditions and evader policies.}
\label{tab:table_cot_1}
\end{table}

The application of the Chain of Thought approach demonstrates a significant improvement in the generalization of spacecraft piloting techniques for the Pursuer-Evader problem as well as guiding the model to achieve a 0\% failure rate in execution. Results in Table \ref{tab:table_cot_1} indicate a consistent achievement of target proximity within 25 meters across varied scenarios, highlighting the approach's effectiveness in enhancing model precision and adaptability in complex rendezvous tasks. It is evident that this advancement notably enhances the model reasoning for its spacecraft operating capabilities.

\subsection{Fine-tuning}
One notable characteristic of control tasks requiring real-time responses is the imperative for quick reactions. OpenAI's ChatGPT completion API incorporates a fine-tuning feature \footnote{GPT-3.5 Turbo fine-tuning and API updates: \url{https://openai.com/blog/gpt-3-5-turbo-fine-tuning-and-api-updates}} for various specified purposes:
\begin{itemize}
    \item Higher quality results than prompting
    \item Ability to train on more examples than can fit in a prompt
    \item Token savings due to shorter prompts
    \item Lower latency requests
\end{itemize}
Given the real-time nature of the KSPDG problem, it is crucial to minimize latency in requests, thereby achieving a more rapid and efficient response to a range of events some of which cannot be predicted. These events may include collision avoidance, orbital adjustments, evasive maneuvering and response to unforeseen obstacles, among other challenges commonly observed in space-related scenarios.

Facing the KSPDG challenge, we rapidly noticed the scarcity of available data for training purposes. Consequently, we developed a script with the capability of recording human gameplays in KSP and turn them into a seq data, in order to train the model to imitate human-like approaches for spacecraft maneuvering. The training data for fine tuning consists of observations and thrust actions (keyboard strokes) which are recorded at regular time intervals during the human gameplay. Thrust actions are logged as discrete labels, relative to the spacecraft's reference frame (e.g. "backward", "forward", "left", "right", "down", "top"). The recording interval was set at half a second, being coherent with the latency presented in Table~\ref{tab:fine_tune_1}

This approach requires the agent to operate within the spacecraft's reference frame, which adds to the model's reasoning. However, it also allows for optimal utilization of the LLM's capabilities.

To evaluate the performance of fine tuning we run the following experiments sequentially, using GPT-3.5-turbo-1106\footnote{As of February 2024, alternative models available with fine tuning capabilities include gpt-3.5-turbo-0613, babbage-002, davinci-002, and gpt-4-0613. However, these demonstrated inferior performance in comparison.} as LLM baseline:
\begin{itemize}
    \item Baseline LLM: This is the standard "gpt-3.5-turbo" model used without any fine-tuning. It is programmed with a generalized system prompt designed for the Pursuer-Evader (PE) scenarios within the KSPDG challenge (See Fig. \ref{fig:system_prompt-fine-tuning}). This prompt encompasses the key aspects of rendezvous missions.
    \item Fine tuning: Fine tuning with a single human gameplay log and default hyperparameters. This file is composed of 314 user-assistant message pairs, each encompassing both observations (user prompt) and corresponding throttle response actions (assistant).
    \item + hyperparameter tuning: After adjusting the values of GPT-3.5 fine tune hyperparameters \cite{openai2023hyperparameters}. We notice that GPT-3.5 tended to select very high learning rate multipliers by default, which caused the optimizer to converge quickly without sufficient exploration. Setting this hyperparameter to 0.2 resulted in better LLM performance.
    \item + system prompt: After adding system prompt to the training data. It is crucial to employ a proper system prompt, maintaining a balance between providing clear and concise instructions and not constraining the model's conduct. This system prompt can be read in Figure \ref{fig:system_prompt-fine-tuning}.
    \item + two train gameplays: After adding an additional human gameplay in the training data. The objective of this experiment is to investigate the rate at which the model acquires knowledge given more data. Consequently, this facilitates an analysis of the availability and potential effectiveness of this methodology. In contrast with the single file approach, this strategy is composed of 647 action messages.
\end{itemize}

\begin{figure}[h]
\centering
\fbox{
\begin{minipage}{.9\textwidth}
\texttt{You operate as an autonomous agent controlling a pursuit spacecraft. Your goal is to apply throttles to capture the evader given the positions and velocities of the pursuer and evader and considering that maximum throttle is 1. }\textbf{\texttt{As a hint you should accelerate in the direction of the relative position to the evader.}}\texttt{ Do not show neither reasoning nor calculations. After reasoning call the perform\_action function.}
\end{minipage}
}
\caption{System prompt used in the fine tuning strategy.}
\label{fig:system_prompt-fine-tuning}
\end{figure}

\begin{table}[h]
\centering
\begin{tabular}{l|ccc}
\hline
\textbf{Method} & \textbf{Best Latency} & \textbf{Average Latency} & \textbf{Standard Deviation} \\ \hline
baseline LLM & 759.69 & 840.42 & 83.85 \\
simple fine-tuning & 977.32	& 987.43 & \textbf{15.08} \\
+ hyperparameter tuning & 749.02 & 831.30 & 49.02 \\
+ system prompt & 684.13 & 753.52 & 51.75 \\
+ two train gameplays & \textbf{468.98} & \textbf{557.49} & 89.54 \\ \hline
\end{tabular}
\caption{Latency of responses for each experiment in milliseconds}
\label{tab:fine_tune_1}
\end{table}

\begin{table}[h]
\centering
\begin{tabular}{l|ccc}
\hline
\textbf{Method} & \textbf{Best Distance (m)} & \textbf{Average Distance (m)} & \textbf{Failure Rate} \\ \hline
Naive & 225.0 & 225.0 & - \\
PPO & >1643 & 2346 & - \\
iLQGames & >53.31 & 60.99 & - \\
Lambert-MPC & >18.24 & \textbf{47.94} & - \\
\hline
baseline LLM & 178.11 & 200.10 & 36.8\% \\
simple fine-tuning & 263.55 & 265.89 & \textbf{0.0\%} \\
+ hyperparameter tuning & 188.90 & 202.08 & 0.1\% \\
+ system prompt & 197.41 & 214.87 & \textbf{0.0\%} \\
+ two train gameplays & \textbf{132.09} & \textbf{159.78} & 0.2\% \\ \hline
\end{tabular}
\caption{Performance of LLM for each fine-tuning technique in meters}
\label{tab:fine_tune_2}
\end{table}

We notice that fine tuning significantly reduces GPT-3.5 response latency as shown in Table~\ref{tab:fine_tune_1}. This is due to the reduced number of output tokens the fine tuning models generate as well as to the large number of failures observed in the LLM baseline as shown in Table~\ref{tab:table_1_label} (most of the failures are function call invocation errors). One particular detail that can be inferred from both Table \ref{tab:fine_tune_1} and Table \ref{tab:fine_tune_2} is that, in contrast to few-shot prompting, fine-tuning yields better results after a higher number of adjustments.

Interestingly, while the OpenAI fine-tuning API requires customization, it offers limited tools to achieve this. Hence, the effectiveness of the training relies heavily on the quantity and quality of data, as well as some of the adjustments previously mentioned. As shown in Table~\ref{tab:table_1_label}, it can be seen that the failure rate of the GPT-3.5 model in this task exceeds tolerable thresholds. In contrast, this problem is completely eradicated in all of the fine-tuned models, resulting in a failure rate close to 0\%. Table~\ref{tab:fine_tune_2} presents a distribution similar to that of Table~\ref{tab:fine_tune_1}. The best distance is only improved by adding all the different increments. In that specific case, the final score surpasses any other naive or baseline technique.



\begin{figure} [htb]
\centering
\begin{subfigure}{.5\textwidth}
  \centering
  \includegraphics[scale=.35]{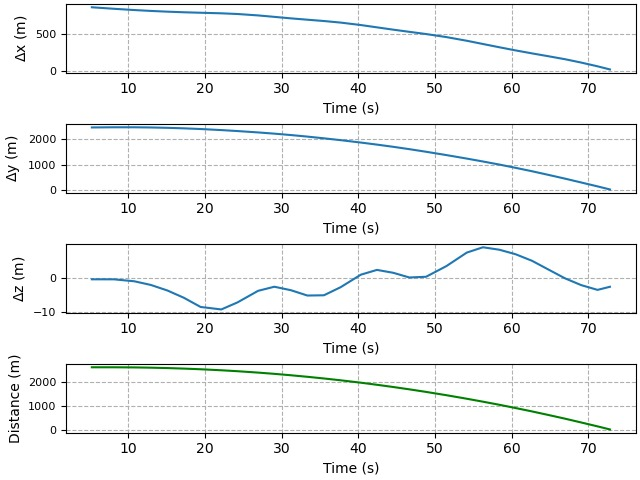}
  \caption{Evolution of differences in the position and distance metric.}
  \label{fig:2d}
\end{subfigure}%
\begin{subfigure}{.5\textwidth}
  \centering
  \includegraphics[scale=.4]{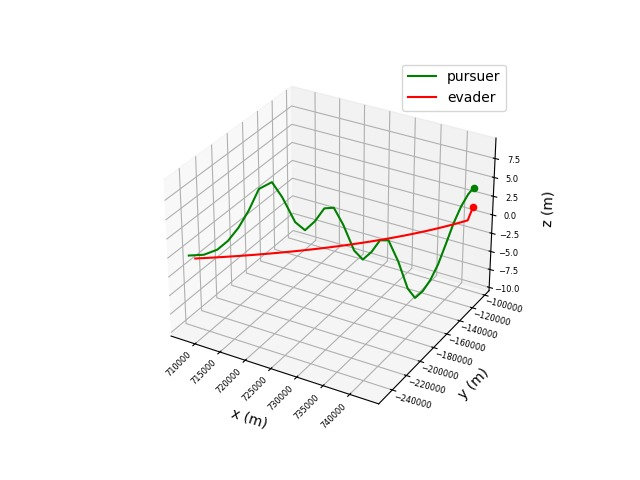}
  \caption{3D trajectory plot plot}
  \label{fig:3d}
\end{subfigure}
\caption{Behavior of the fine-tuned LLM agent in a Pursuer-Evader scenario. The 2D plot (a) shows the relative position and distance between the pursuer and evader over time, while the 3D plot (b) depicts their trajectories in space.}
\label{fig:trajectories}
\end{figure}

Figure \ref{fig:trajectories} depicts the model's behavior over time. The agent demonstrates its ability to quickly close the distance between the pursuer and evader, as evidenced by the sharp decrease in distance within the first 20 seconds of the mission. However, once the pursuer reaches a certain proximity to the evader, the agent struggles to maintain a consistent close distance, as shown by the oscillations in the relative position and distance plots. This suggests that while the fine-tuning strategy enables the LLM to effectively guide the pursuer towards the evader, it may lack the necessary fine-grained control to maintain a stable, close pursuit once the initial approach is complete. Further refinements to the fine-tuning process, such as incorporating more diverse training data or adjusting the model's hyperparameters, could potentially address this limitation and improve the agent's ability to sustain a close pursuit throughout the mission.

\section{Conclusions and future work}
In this work, we have studied the potential of Large Language Models (LLMs) as autonomous agents for spacecraft control in the context of the Kerbal Space Program Differential Games (KSPDG) challenge. By leveraging the power of OpenAI's GPT-3.5 and developing some strategies on top of it such as prompt engineering, few-shot prompting, and fine-tuning it with human gameplay data, we developed an LLM-based agent capable of performing spacecraft maneuvers in pursuit-evasion scenarios. Our agent achieved competitive results, ranking 2nd in the KSPDG challenge, and showcased the effectiveness of LLMs in solving control problems in the space domain.

The success of our LLM-based approach highlights the limitations of traditional Reinforcement Learning (RL) methods in space applications. RL algorithms often require a large number of simulations and a well-defined reward function to learn effective control policies. However, in the space domain, simulations are often scarce, and defining a suitable reward function can be challenging. LLMs, on the other hand, leverage their vast pre-trained knowledge base and can be fine-tuned to specific tasks with relatively small amounts of data. This makes LLMs particularly well-suited for space applications, where data and simulations are limited.

It is important to note that the application of LLMs in space is not limited to digital assistants and language-based tasks. As demonstrated in this work, LLMs can be used as autonomous agents capable of making decisions and taking actions based on numerical data and complex environments. This opens up a whole new realm of possibilities for AI in space, beyond the traditional language-centric applications. We can expect to see them applied to a wide range of problems in the space domain, from autonomous planetary landing and sensor tasking to spacecraft maneuvering and mission planning. 

However, the integration of LLMs into critical space missions also poses significant challenges. Given the current limitations of LLMs, such as their propensity for hallucinations and the difficulty in interpreting their decision-making process, it is crucial to develop rigorous testing procedures to ensure the reliability and safety of LLM-based systems in space applications. As LLMs continue to improve in terms of accuracy and robustness, we can expect to see a gradual adoption of these models in increasingly critical space missions.

Based on the findings and insights gained from this research, we propose the following directions for future work:
\begin{itemize}
    \item Investigate the performance of various LLMs, including both proprietary (e.g., GPT, Anthropic's Claude) and open-source models (e.g., Llama, Mistral), in the context of the KSPDG challenge. This study would also take into account the computational requirements and model sizes, since it is crucial to ensure that the LLMs can operate within the constraints of the available hardware in space.
    
    \item Explore the application of Large Multimodal Models (LMMs) in the KSPDG environment by incorporating visual inputs, such as screenshots of the gameplay, alongside telemetry data. This approach could lead to a more generalizable solution for non-cooperative space missions, where accurate observations of the evader spacecraft may not be available..
    
    \item Investigate the use of LLMs as code generators to create autonomous agents for spacecraft control. Instead of using LLMs directly as the policy in a reinforcement learning loop, we can use LLMs to generate the code for the agent itself. By iterating on the generated code based on the feedback from simulations, LLMs could potentially create high-performing agents in just a few iterations, leveraging their vast knowledge and coding capabilities. This approach would result in a standalone agent that does not require real-time execution of the LLM, making it more suitable for on-board deployment, and less token-consuming.
    
    \item Study the scalability of fine-tuning LLMs with larger and more diverse datasets. In this work, we used a limited number of human gameplay logs for fine-tuning due to the high cost associated with proprietary models like OpenAI's GPT-3.5. Future research should investigate the impact of fine-tuning with datasets that encompass a wide range of missions, initial conditions, evader policies, and pursuer positions. This study could be conducted using open-source models to mitigate the cost constraints and provide a more comprehensive understanding of how fine-tuning scales with data size and diversity.
    
    \item Extend the application of LLMs to other areas in the space domain where reinforcement learning has struggled to create effective autonomous agents, such as sensor tasking, planetary landing, and autonomous navigation.
\end{itemize}

\section*{Acknowledgments}
Authors would like to acknowledge Dr. Ross Allen and the rest of the team behind the Kerbal Space Program Differential Game (KSDPG) challenge at MIT Lincoln Laboratories, for the creation of the tool that triggered the development of this research and the support received during the challenge.

\bibliographystyle{unsrt}  
\bibliography{references}

\end{document}